\documentclass{iaus}
\usepackage{graphicx}

\title[Stellar surface rotation] 
{Investigating stellar surface rotation using observations of starspots}

\author[Heidi Korhonen]   
{Heidi Korhonen$^{1,2,3}$}

\affiliation{$^1$Niels Bohr Institute, University of Copenhagen, Juliane 
Maries Vej 30, DK-2100 Copenhagen, Denmark
 email: {\tt heidi.korhonen@nbi.ku.dk} \\[\affilskip]
$^2$Finnish Centre for Astronomy with ESO (FINCA), University of Turku, 
V{\"a}is{\"a}l{\"a}ntie 20, FI-21500 Piikki{\"o}, Finland\\[\affilskip]
$^3$Centre for Star and Planet Formation, Natural History Museum of Denmark, 
University of Copenhagen, {\O}ster Voldgade 5-7, DK-1350, Copenhagen, Denmark\\}

\pubyear{2008}
\volume{xxx}  
\pagerange{119--126}
\jname{Title of your IAU Symposium}
\editors{A.C. Editor, B.D. Editor \& C.E. Editor, eds.}
\begin{document}

\maketitle

\begin{abstract}
Rapid rotation enhances the dynamo operating in stars, and thus also introduces
significantly stronger magnetic activity than is seen in slower rotators. Many 
young cool stars still have the rapid, primordial rotation rates induced by the 
interstellar molecular cloud from which they were formed. Also older stars in 
close binary systems are often rapid rotators. These types of stars can show 
strong magnetic activity and large starspots. In the case of large starspots 
which cause observable changes in the brightness of the star, and even in the 
shapes of the spectral line profiles, one can get information on the rotation 
of the star. At times even information on the spot rotation at different 
stellar latitudes can be obtained, similarly to the solar surface differential 
rotation measurements using magnetic features as tracers. Here, I will review 
investigations of stellar rotation based on starspots. I will discuss what we 
can obtain from ground-based photometry and how that improves with the 
uninterrupted, high precision, observations from space. The emphasis will be on
how starspots, and even stellar surface differential rotation, can be studied 
using high resolution spectra.

\end{abstract}

\firstsection 

\section{Introduction}

It is widely accepted that the global behaviour of the solar magnetic field
can be explained by a dynamo action which is due to interaction between
magnetic fields and fluid motions. The Sun is thought to have an 
$\alpha\Omega$-type dynamo, in which the poloidal field is created from the 
toroidal one by helical turbulence ($\alpha$-effect), and the toroidal field 
is obtained by shearing the already existing poloidal field by differential 
rotation ($\Omega$-effect). This kind of dynamo is also thought to work in 
other main-sequence stars with similar internal structure as the Sun has, i.e.,
stars with masses of $\sim$0.4--2.0~M$_{\odot}$.

Stellar rotation has a major impact on the over-all efficiency of the dynamo 
action, and thus on the level of observed magnetic activity. The relationship 
between rotation and activity was first studied in detail by \cite[Pallavicini 
\etal\ (1981)]{Pallavicini81} who investigated the correlation between the 
X-ray luminosity and projected rotation velocity, $v\sin i$. Since then several
studies have shown that the magnetic activity increases with increasing 
rotation rate, until finding a rotation rate after which no increase, and even 
maybe a small decrease, occurs (e.g., \cite[Micela \etal\ 1985]{Micela85}; 
\cite[Pizzolato \etal\ 2003]{Pizzolato03}; \cite[Wright \etal\ 
2011]{Wright11}). This so-called saturation limit of the magnetic activity is 
reached at certain spectral type dependent rotation period, which increases 
toward later spectral types (\cite[Pizzolato \etal\ 2003]{Pizzolato03}).

There is also evidence that the stellar cycle length correlates to some extend 
with the rotation rate. Faster rotation tends to create shorter activity cycles
(e.g., \cite[Noyes \etal\ 1984]{Noyes84}; \cite[Saar \& Brandenburg 1999]
{Saar99}; \cite[Ol{\'a}h \etal\ 2000]{Olah00}). In a diagram where the cycle 
frequency ($\omega_{\rm cycl}$) is plotted against the rotational frequency 
$\Omega$ the stars also seem to occupy three different regions: so-called 
inactive, active, and super-active regions (e.g., \cite[Saar \& 
Baliunas 1992]{Saar92}; \cite[Saar \& Brandenburg 1999]{Saar99}).

The studies of open clusters with known ages show that young clusters, with age 
of some tens of Myrs, have many rapidly rotating cool stars. With increasing 
age the amount of cool rapidly rotating stars decreases, and around 600 Myrs 
the stars with similar internal structure to the Sun's have slowed down to 
relation that is dependent on stellar mass (see e.g., 
\cite[Barnes 2003]{Barnes03}; \cite[Meibom \etal\ 2011]{Meibom11}). The 
observed spin-down of the stars over their lifetimes can be attributed to the 
magnetic breaking, which is driven by mass-loss through a magnetised stellar 
wind (e.g., \cite[Skumanich 1972]{Skumanich72}). Therefore, many young cool 
stars (spectral classes G--M) are rapid rotators. The rapid rotation introduces
significantly stronger magnetic activity than is seen in their older main 
sequence counterparts, like the Sun. On the other hand, the Sun was most likely
in its youth a very active stars. All the energetic events caused by the 
magnetic activity can heat the possible planet forming disc around the star and
have an impact on its composition, thus also affecting the possible planet and 
planetary system formation process. The phenomena caused by stellar activity 
can also have similar effects on stellar brightness and radial velocity as 
orbiting planets, making it at times difficult to distinguish between planets 
and activity signatures (see, e.g., \cite[Queloz \etal\ 2001]{Queloz01}). 

It is clear that stellar rotation has a significant effect on magnetic activity
of stars. In this review I will discuss how to measure stellar rotation using 
starspots, and especially concentrating on detailed studies of stellar surface 
differential rotation, which is a crucial parameter in the solar and stellar 
dynamos.

\section{Methods for studying stellar rotation using starspots}

Stars are point sources and studying their surface features is very demanding. 
For a long time it was not possible to obtain direct, spatially resolved, images
of the stellar surface, except in some very rare cases of near-by giant and 
supergiant stars, like Betelgeuse (\cite[Gilliland \& Dupree 1996]
{Gilliland96}). During the last years a breakthrough using long baseline 
infrared interferometers has occurred. Images with milli-arcsecond resolution 
can now be produced, and a variety of targets have been imaged with astonishing
result, e.g., bulging stars rotating near their critical limit (\cite[Monnier 
\etal\ 2007]{Monnier07}), and images of the transiting disk in the $\epsilon$ 
Aurigae system (\cite[Kloppenborg \etal\ 2010]{Kloppenborg10}). Infrared 
interferometric imaging has produced amazing results of stellar surfaces and 
the time is drawing near when even temperature spots on cool stars can be 
imaged. In addition, helioseismology has given us an unprecedened view of the 
solar rotation, both at the surface and in the interior, but obtaining similar 
results for stars is still very much 'work in progress'. 

The main methods for studying stellar rotation are similar to the ones used 
originally on the Sun, i.e., tracing the movement of magnetic features, mainly 
starspots. This can be done using photometry and high resolution spectroscopy 
through Doppler imaging techniques. The rest of the article is focused on these 
two methods. Still, it has to be remarked that stellar rotation has also been 
studied using spectral line profile shapes (e.g., \cite[Reiners \& Schmitt 
2003]{Reiners_Schmitt03}) and Ca II H\&K emission variations (e.g., 
\cite[Donahue \etal\ 1996]{Donahue96}; \cite[Katsova \etal\ 2010]{Katsova10}).

\subsubsection{Photometry}

In many active stars the starspots are so large that they cause brightness 
variations which can be few tens of percent from the mean light level, thus 
easily observable even with small ground-based telescopes (see, e.g., 
\cite[Kron 1947]{Kron47}). In comparison, biggest sunspots would only cause 
$\sim$0.01\% decrease in the solar brightens and would require extremely 
precise instruments to detect. 

Numerous studies of stellar rotation on active stars have been carried out 
based on photometry. Long-term monitoring campaigns have been carried out by 
several groups resulting in numerous papers on stellar rotation (e.g., 
\cite[Messina \& Guinan 2003]{Messina03}; \cite[Strassmeier \etal\ 1997]
{Strassmeier97}). Most of the campaigns are primarily used for studying 
long-term stellar cycles, but also the rotation periods and their possible 
variations have been investigated. 

\begin{figure}[t]
\begin{center}
 \includegraphics[width=3.4in]{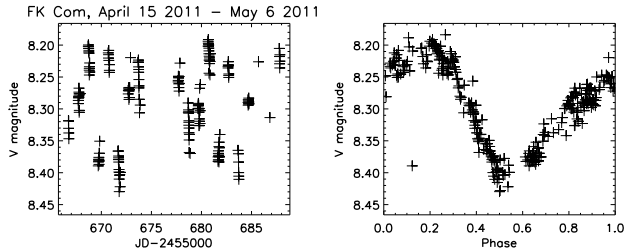} 
 \caption{Example of ground-based observations of an active star FK Com. The 
plot gives the V magnitudes plotted against the Julian Date (left panel) and 
against the phase (right panel). The observations have been obtained in 
April--May 2011 using automatic photometric telescopes in Arizona. More 
information on the telescopes is given by \cite[Strassmeier \etal\ 
(1997)]{Strassmeier97}.}
   \label{fig1}
\end{center}
\end{figure}

In general, the observations are carried out by automatic telescopes and 
typically couple of observations per night are obtained during as many nights 
as possible within the observing season. Depending on the exact stellar 
rotation period this results in a time series from which the period might be 
difficult to determine accurately. Fig.\ \ref{fig1} gives an example of a 
ground-based light-curve of an active star, in this case the single yellow 
giant FK~Com. These observations actually present a good phase coverage 
light-curve obtained during a dedicated campaign. The typical dataset has much 
less observations within similar time period. 

\begin{figure}[t]
\begin{center}
 \includegraphics[width=4in]{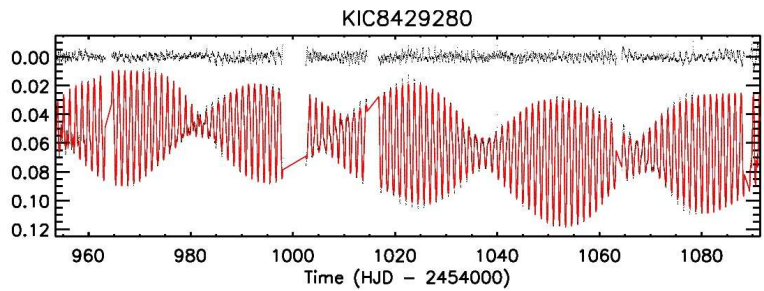} 
 \caption{Example of a high precision continuous space-based observations of a 
young active star. The observations have been obtained using Kepler and have 
been published by \cite[Frasca \etal\ (2011)]{Frasca11}.}
   \label{fig2}
\end{center}
\end{figure}

If the period is not straight forward to determine from the often sparse 
ground-based observations, it is one of the easiest stellar properties to 
obtain from continuous space-based observations. With the recent launch of 
several photometric space missions, the high accuracy, high cadence, stellar 
light-curves are now also revolutionising the active star research. These 
missions include the Canadian micro-satellite MOST (for results on active stars
see, e.g., \cite[Rucinski \etal\ 2004]{Rucinski04} and \cite[Siwak \etal\ 2011]
{Siwak11}) and the French CoRoT satellite (e.g., \cite[Lanza \etal\ 2009]
{Lanza09}; \cite[Silva-Valio \etal\ 2010]{SilvaValio10}; \cite[Huber \etal\ 
2010]{Huber10}). The real break-through happened, though, with the launch of 
the NASA's Kepler satellite. Kepler provides unprecedented accuracy 
light-curves of about 150~000 stars, among them several active stars (see 
Fig.~\ref{fig2} and, e.g., \cite[Frasca \etal\ 2011]{Frasca11}). Based on GALEX
data a sample of about 200 active stars have been selected for Kepler 
observations in a Guest Observer programme (\cite[Brown \etal\ 2011]{Brown11}).
The space-based data has shown us that basically every stellar rotation has 
slightly different spot configuration. This casts serious doubts in the old
practice of using ground-based data from several rotations together.

\subsubsection{Doppler imaging}

Doppler imaging is a method that can be used for detailed mapping of the 
stellar surface structure (e.g., \cite[Vogt \etal\ 1987]{Vogt87}; 
\cite[Piskunov \etal\ 1990]{Piskunov90}). Hereby high resolution, high 
signal-to-noise spectra at different rotational phases are used to measure the 
rotationally modulated distortions in the line-profiles. These distortions are 
produced by the inhomogeneous distribution of the observed characteristic, 
e.g., effective temperature, element abundance or magnetic field. Surface maps,
or Doppler images, are constructed by combining all the observations from 
different phases and comparing them with synthetic model line profiles. Doppler 
imaging techniques were first used in the abundance mapping of Ap stars. 
Nowadays, Doppler imaging is more commonly used for temperature mapping of 
rapidly rotating late type stars (e.g., \cite[Korhonen \etal\ 
2007]{Korhonen07}; \cite[Skelly \etal\ 2010]{Skelly10}). 

One has to keep in mind though, that for successful Doppler imaging a priori 
knowledge of the stellar rotation, usually based on earlier photometric 
investigations, is needed. If the rotation period of the Doppler imaging target
is not know, it is very difficult to plan the observations so that a good 
phase coverage needed for Doppler imaging is obtained. Also, if the rotation 
period is not known accurately, comparison of the maps recovered at different 
epochs is not straight forward. The spectra and the variations seen in them can
be used for estimating the rotation period of the target, but still the best 
results are obtained if also long-term photometric monitoring is carried out 
for accurate period determination.

Until autumn 2011 Doppler imaging has been applied to some 70 stars. 
\cite[Strassmeier (2009)]{Strassmeier09} gives a recent review on starspots and
their properties. Here only a short summary on the main results we have learned
about starspots using Doppler imaging is given.

\begin{figure}[t]
\begin{center}
 \includegraphics[width=3.4in]{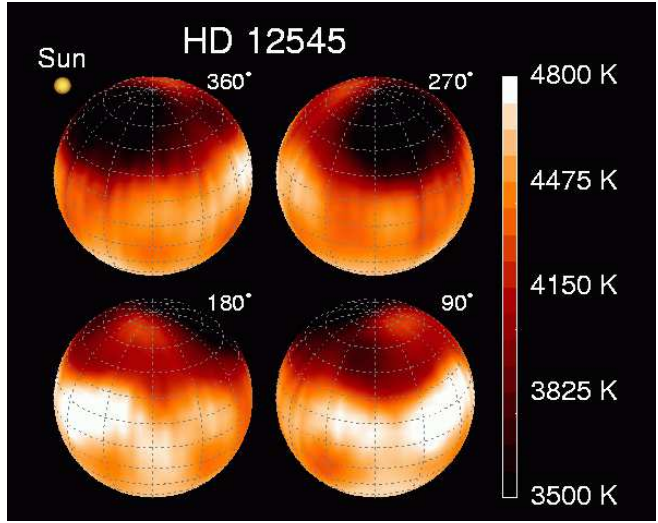} 
 \caption{Doppler imaging results of HD~12545 by \cite[Strassmeier (1999)]
{Strassmeier99}.}
   \label{fig3}
\end{center}
\end{figure}

Due to the enhanced magnetic activity in rapidly rotating cool stars, the spots
are much larger than the spots observed in the Sun. The largest starspot 
recovered with Doppler imaging is on the active RS~CVn binary HD 12545 which, 
in January 1998, had a spot that extended approximately 12$\times$20 solar 
radii (\cite[Strassmeier 1999]{Strassmeier99}). The image of this starspot is 
shown in Fig.\,\ref{fig3}. 

The lifetime of the large starspots can also be much longer than that of the 
sunspots, even years instead of weeks for sunspots (e.g., \cite[Rice \& 
Strassmeier 1996]{Rice96}; \cite[Hussain 2002]{Hussain02}). One has to 
keep in mind though, that the spatial resolution obtained with Doppler imaging 
is not good enough for distinguishing whether the large spots are one single 
spot or a group of spots. If they are groups of spots, the individual spots 
could exhibit much shorter lifetimes than the group itself. Based on numerical 
simulations \cite[I\c{s}{\i}k \etal\ (2007)]{isik07} have shown that in a 
rapidly rotating active star the expected spot lifetime is few months, 
depending on the spot latitude (mid-latitude spots live a shorter time than 
equatorial or high latitude spots) and spot size (larger spots live longer). 
For sub-giants even longer spot lifetimes are obtained.

The latitudes at which starspots often occur are very different from those
for the sunspots. In rapidly rotating late type stars spots can appear at 
very high latitudes, unlike in the Sun. This can be explained by 
the increase in the Coriolis force induced by the rapid rotation (e.g., 
\cite[Sch{\"u}ssler \& Solanki 1992]{Schussler92}; \cite[Granzer \etal\ 
2000]{Granzer00}). Still, these calculations cannot explain the formation of 
the polar caps, i.e., spots that are located at the rotational poles of the 
star, except in very young stars. These polar caps are still often seen in 
the Doppler images of also older late type stars (e.g., \cite[Weber \& 
Strassmeier 2001]{Weber01}).

\section{Surface differential rotation}

Gaseous bodies, like the stars, can rotate differentially, i.e., different 
latitudes can have different rotation rates. In the Sun the rotation velocity 
of the photosphere depends strongly on the latitude; the rotation of the solar 
equator is approximately 30\% shorter than the period at the poles. 
Helioseismological studies show that this latitude dependence persists 
throughout the convection zone (e.g., \cite[Thompson \etal\ 1996]{Thompson96}). 

Differential rotation is one of the main elements in the dynamo models. 
Together with the helical turbulence and meridional flow it is responsible for 
the main features of the solar and stellar magnetic activity (see, e.g., 
\cite[ R{\"u}diger \etal\ 1986]{rued86}; \cite[Brun \& Toomre 2002]{Brun02}). 
Therefore, it is important to also measure differential rotation on other stars
than the Sun.

Usually it is assumed that the differential rotation law of the Sun can be 
generalised to stars, leading the surface rotation law to be expressed by

\begin{equation}
\Omega=\Omega_{\rm eq}+\beta\sin^{2}\psi,
\label{eq:lat}
\end{equation}
where $\psi$ is the latitude, $\Omega_{\rm eq}$ is the equatorial angular
velocity and $\beta$ defines the magnitude of the differential rotation. The 
relative differential rotation coefficient is given by
\begin{equation}
\alpha=\frac{\Omega_{\rm eq}-\Omega_{\rm pol}}{\Omega_{\rm eq}}\ \ {\rm or}\ 
\alpha=\frac{-\beta}{\Omega_{\rm eq}}
\label{eq:alpha}
\end{equation}
where $\Omega_{\rm pol}$ is the polar angular velocity. Note though that the 
exact formulation, symbols and signs, vary from author to author. Therefore, one
has to be careful when comparing different works.

\begin{figure}[t]
\begin{center}
 \includegraphics[width=2.0in]{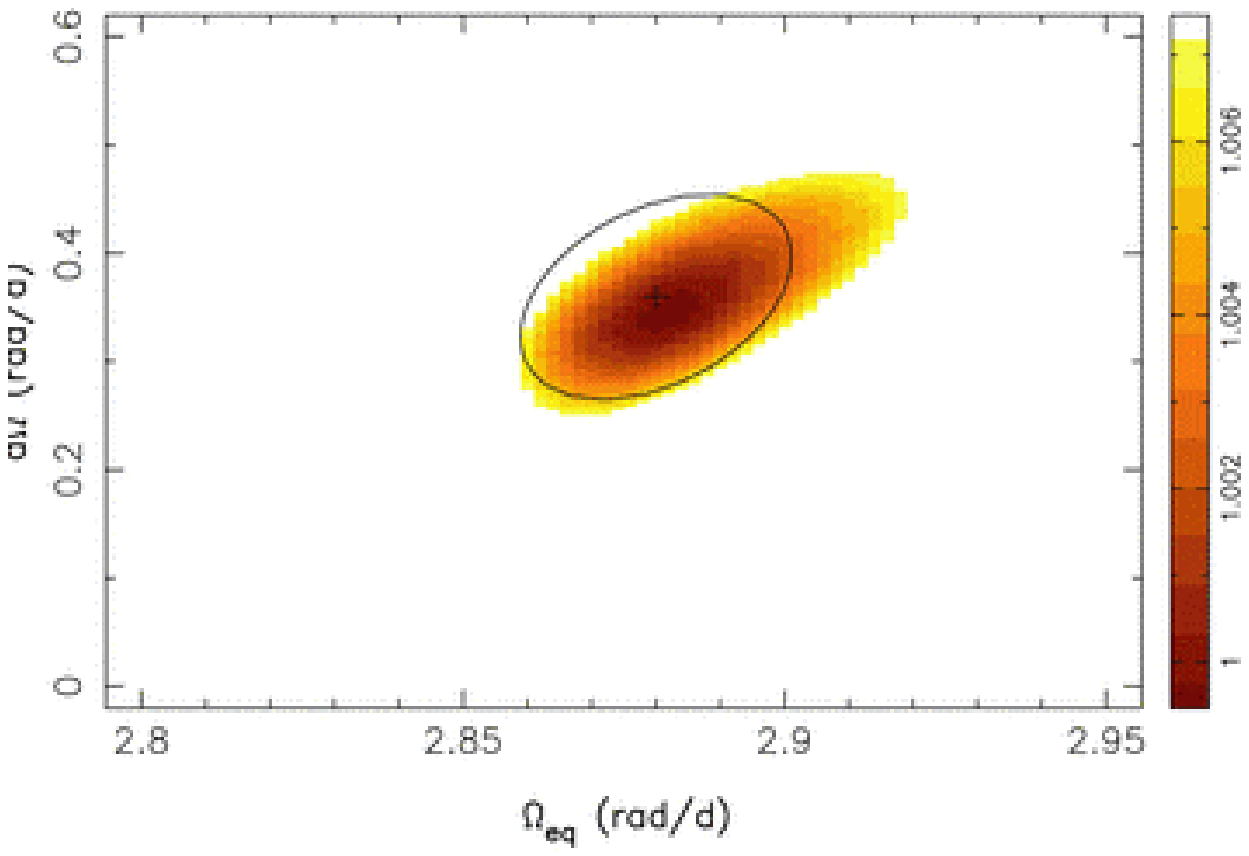} 
 \includegraphics[width=3.0in]{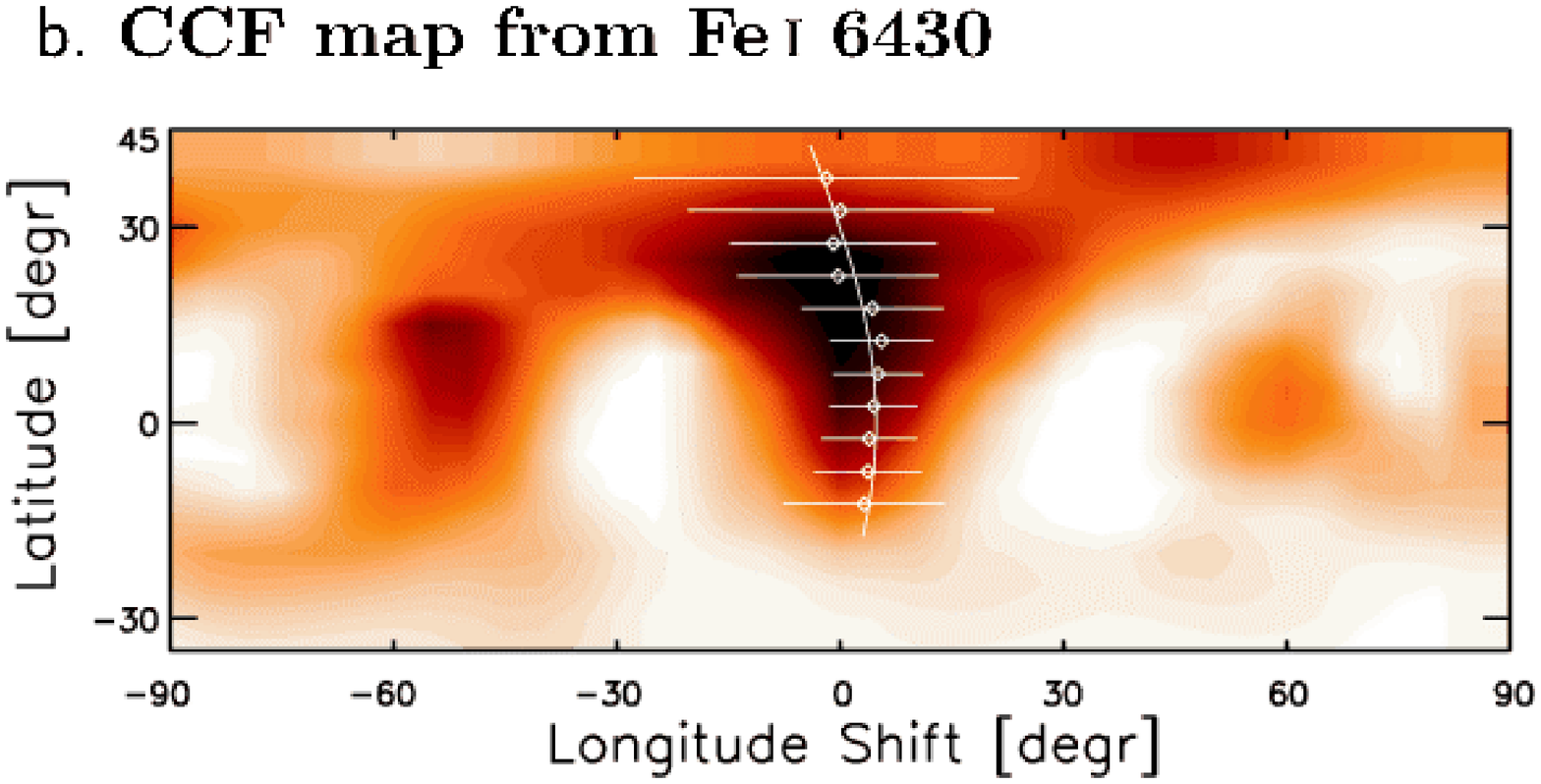} 
 \caption{Examples of surface differential rotation determination using 
$\chi^2$-landscape (left; \cite[Marsden \etal\ 2011]{Marsden11}) and 
cross-correlation (right; \cite[K{\H o}v{\'a}ri \etal\ 2007a]{Kovari07a}) 
methods.}
   \label{fig4}
\end{center}
\end{figure}

Measuring stellar differential rotation is not straight forward. Currently the 
best way to study stellar surface DR in detail, both the strength and the sign,
is by using surface maps obtained with Doppler imaging. One way, similar to 
using sunspots and other magnetic features to study the solar surface 
differential rotation, is to cross-correlate several surface maps obtained at 
different times with Doppler imaging. This allows to investigate the changes in
the locations of the spots and how that depends on the latitude (see, e.g., 
\cite[Barnes \etal\ 2000]{Barnes00}; \cite[Weber \etal\ 2005]{Weber05}; 
\cite[K{\H o}v{\'a}ri \etal\ 2007a]{Kovari07a}). The other often applied 
method is so-called $\chi^2$-landscape technique in which a solar-like surface 
differential law is implemented into the Doppler imaging code, and many maps 
are obtained to study which parameters give the best solution (e.g., 
\cite[Petit \etal\ 2002]{Petit02}; \cite[Marsden \etal\ 2006]{marsd06}; 
\cite[Dunstone \etal\ 2008]{Dunstone08}). For examples of both methods, look at 
Fig.~\ref{fig4}. Stellar surface differential rotation has also been studied by 
combining spot latitude information from Doppler images and spot rotation 
period from contemporaneous photometry 
(\cite[Korhonen \etal\ 2007]{Korhonen07}).

\subsection{Surface differential rotation with spectral type and rotation 
period}

\cite[Donahue \etal\ (1996)]{Donahue96} investigated the rotation periods and 
the change in the rotation period, i.e., magnitude of the surface differential 
rotation, in 36 stars from Ca II H\&K S-index measurements. They found that 
there was a power-law correspondence between these values, where $\Delta\Omega$
is proportional to the mean seasonal rotation, $\Omega$, with the power of 0.7,
independent of mass. \cite[Barnes \etal\ (2005)]{Barnes05} collected 
measurements from different sources and methods, and obtained results which 
also imply power-law correlation between the rotation and surface differential 
rotation, but with a power of only 0.15. On the other hand, recent 
investigation by \cite[Saar (2011)]{Saar11} gives relation $\Delta\Omega\propto 
\Omega^{0.68}$, i.e. similar to the \cite[Donahue \etal\ (1996)]{Donahue96} 
results. It is clear that selecting homogeneous sample for these studies is 
difficult. Additionally, the surface differential rotation estimates used in 
some of these studies come from different methods and exhibit different 
systematic effects, therefore making comparison difficult.

The theoretical calculations by \cite[Kitchatinov \& R{\"u}diger 
(1999)]{Kitchatinov99} show that the absolute value of the surface differential
rotation decreases initially as the rotation period decreases from the solar 
value, but changes to a slight increase for periods of few days. They also 
predict that the differential rotation for the giant stars is larger than that 
for the dwarfs. Surface differential rotation measurements of 10 young G2--M2 
dwarfs were obtained using Doppler imaging by \cite[Barnes \etal\ 
(2005)]{Barnes05}. These measurements show an increase in the magnitude of 
differential rotation towards earlier spectral types, which is consistent with 
the theoretical calculations. Similar results have also been obtained by 
\cite[Saar (2011)]{Saar11}.

\cite[K{\"u}ker \etal\ (2011)]{Kuker11} carried out an theoretical study of 
rotation of G dwarfs. They computed model convection zones of different depth 
and investigated their large-scale gas motions, i.e. rotation and meridional 
flow, and compared the results to observations. Their calculations could easily
produce the rotation laws of the slowly rotating Sun and several rapidly 
rotating G dwarfs. Their calculations failed to explain the extreme surface 
shear of HD 171488 (reported, e.g., by \cite[Marsden \etal\ 2005]{Marsden05} and
\cite[Jeffers \& Donati 2009]{Jeffers09}), except when using an artificially 
shallow convection zone. This high-lights the fact that even though theory and 
observations at times encouragingly agree, there are still many unexplained 
features in the solar and stellar magnetic activity.

\subsection{Anti-solar differential rotation}

In general, models for global circulation in outer stellar convection zones 
predict solar-type differential rotation, where the equator is rotating faster 
than the poles. However, \cite[Kitchatinov \& R{\"u}diger 
(2004)]{Kitchatinov04} have shown that anti-solar differential rotation could 
arise as a result of intensive meridional circulation.

Anti-solar differential rotation, where the polar regions rotate faster than 
the equator, has been suggested by observations of several active stars (e.g., 
\cite[Vogt \etal\ 1999]{Vogt99}; \cite[Weber 2007]{Weber07}; 
\cite[K{\H o}v{\'a}ri \etal\ 2007]{Kovari07b}). \cite[K{\H o}v{\'a}ri \etal\ 
(2007)]{Kovari07b} investigated surface flow patterns on $\sigma$~Gem from a 
series of observation spanning 3.6 consecutive rotation cycles, and found an 
anti-solar differential rotation with the surface shear of -0.022$\pm$0.006 
(approximately 1\% of the shear in the Sun, but of opposite sign). 
Additionally, they found evidence of a poleward migration trend of spots with 
an average velocity of $\sim$300 m/s. The strong meridional flow hinted at 
$\sigma$~Gem would support the hypothesis of \cite[Kitchatinov \& R{\"u}diger 
(2004)]{Kitchatinov04}, which attributes the anti-solar differential rotation 
to strong meridional circulation. Similar trend is seen by \cite[Weber (2007)]
{Weber07} in the investigation of several active stars. One has to note though,
that these meridional flow measurements can arise from artefacts in maps and 
have to be confirmed with data from several epochs.

\subsection{Temporal variations in surface differential rotation}

Intriguingly, temporal evolution of the surface differential rotation has been 
reported for two young single K stars, AB~Dor and LQ~Hya (see, e.g., 
\cite[Donati \etal\ 2003]{Donati03}; \cite[Jeffers \etal\ 2007]{Jeffers07}. And
in this case the question is not about small-scale variations, like seen on the
Sun, but changes that are as large as 50\% of the mean $\Delta\Omega$. 
\cite[Donati \etal\ (2003)]{Donati03} hypothesise that these temporal 
variations could be caused by the stellar magnetic cycle converting 
periodically kinetic energy within the convective zone into large-scale 
magnetic fields and vice versa, as originally proposed by \cite[Applegate
(1992)]{Applegate92}. They also remark that a definite demonstration of the 
temporal variation would require  monitoring a few stars for a long time, and 
seeing both the differential rotation parameters and the activity proxies 
showing the same cyclical variations, or at least to exhibit strongly 
correlated fluctuations in the case of non-cyclic behaviour.

\subsection{Cautionary remark on surface differential rotation studies}

A recent study by \cite[Korhonen \& Elstner (2011)]{Korhonen11} used snapshots 
from dynamo models to investigate how well the cross-correlation method 
reproduces the latitudinal rotation rates used in the dynamo calculations. 
Their investigation showed that the cross-correlation method works well when 
the time difference between the maps is appropriate for recovering the surface 
differential rotation, and, importantly, if small-scale fields were included in 
the dynamo calculations. Using only the large-scale dynamo field the solution 
was dominated by the geometry of the dynamo field and the input rotation law 
was not recovered from the snapshots. Actually, the results in these cases 
showed much smaller surface differential rotation, than what was used in 
calculating the dynamo models. On the other hand, with additional injection of 
small-scale fields the input surface rotation law was well recovered. 

This rises the question whether the large starspot seen in active stars can 
actually be created by small scale fields. If they are manifestations of the 
large-scale dynamo field, then according to the study by \cite[Korhonen \& 
Elstner (2011)]{Korhonen11} we would not even expect them to follow the surface
differential rotation. Their results also show that the exact latitude 
dependence of the rotation changes during the stellar cycle. Which could 
explain the temporal variation of surface differential rotation discussed above.

\section{Concluding remarks}

On the whole photometry is the easiest and least time consuming way of carrying
out stellar rotation studies. Long time series of observations can easily be 
obtained for even relatively faint targets. Still, the usual cadence for this 
kind of observations is few observations per night -- at most. This usually 
results in data from several different stellar rotation been used together, and
also in difficulties to pin-point the stellar rotation period accurately. 
Continuous high cadence space-based observations have shown us that the 
light-curves of active stars change from rotation to rotation. Thus the 
ground-based observations usually give us only an average spot configuration. 
The high precision light-curves from space missions, like CoRoT and Kepler, on 
the other hand provide us a unique opportunity to investigate stellar rotation 
and starspots with high temporal resolution. Actually, measuring the stellar 
rotation period from ground-based observations is often challenging, whereas 
from CoRoT and Kepler light-curves that is one of the easiest properties to 
determine.

The strength of the ground-based observations lies in investigating stellar 
cycles. Such long-term time series are virtually impossible to obtain from 
space due to the limited mission lifetimes. Also, instrumental artefacts, like 
trends, are often a problem in space-based instruments, again hampering 
accurate studies of stellar cycles. Whereas obtaining rotational periods with 
high precision can uniquely be done using space-based continuous photometry.

Surface differential rotation can be estimated using photometric observations. 
But the investigations suffer from the fact that the information on the spot 
latitudes is usually impossible to obtain, and thus no information on the sign 
of the differential rotation can be obtained. Also, the latitude range of the 
spots is unknown, therefore only a lower estimate of the magnitude of the 
differential rotation can be obtained. Although, from high precision 
space-based photometry it might be possible to obtain also the latitude 
information.

For obtaining the magnitude and sign of the surface differential rotation 
detailed surface maps are needed. Still, also the studies using Doppler images 
are not without problems. All the methods based on Doppler imaging suffer from 
the often restricted latitude range the starspots occur on, and from the 
possible artefacts in the maps. These artefacts can for example rise from 
incorrect modelling of the spectral line profiles. For example incorrect 
modelling of the line core would produce a signal which is always visible around
zero velocity, i.e., would result in a polar spot. Additionally, in 
$\chi^2$-landscape technique a predefined solar-type rotation law is assumed 
and thus no other latitude dependence can be recovered.  On the other hand, in 
cross-correlation the time difference between the maps is crucial, with too 
small difference no change has time to occur, and with too long difference spot
evolution due to flux emergence and disappearance can occur. Therefore, it is 
very demanding to obtain reliable measurements of the stellar surface 
differential rotation.

Even though, these concluding remarks offer more concerns than definite 
answers, it does not mean that the future of studying stellar rotation is bleak.
Doppler imaging keeps on offering us an intriguing picture of stellar surface 
features and the space-based photometry is opening a real golden era for 
stellar rotation studies. In the future also asteroseismology and 
infrared/optical interferometry have a role to play in the new discoveries.

{\bf Acknowledgments}
The author acknowledges the support from the European Commission under the Marie
Curie IEF Programme in FP7.

\begin{discussion}

\end{discussion}

\end{document}